\documentclass[]{mn2e}
\usepackage[dvips]{graphicx}
\title[Observing the double emission Line feature from 
primordial cloud cores]
{ On the possibility of observing the double emission line feature of
H$_2$ and HD from primordial molecular cloud cores}
\author[H.Kamaya and J.Silk]
{Hideyuki Kamaya$^{1}$\thanks{Email:kamaya@kusastro.kyoto-u.ac.jp}
and Joseph Silk$^{2}$ \\
$^{1}$ Department of Astronomy, School of Science,  
Kyoto University, Kyoto, 606-8502, Japan \\
$^{2}$ Astrophysics, Department of Physics, Denys Wilkinson Building, 
University of Oxford, Oxford OX1 3RH 
}

\begin{document}
\maketitle 
\begin{abstract}

We study the prospects for observing H$_2$ and HD emission during the
assembly of primordial molecular cloud cores.  The primordial
molecular cloud cores, which resemble those at the present epoch, can
emerge around $1+z \sim 20$ according to recent numerical simulations.
A core typically contracts to form the first generation of stars and
the contracting core emits H$_2$ and HD line radiation. These lines
show a double peak feature.  The higher peak is the H$_2$ line of the
$J=2-0$ (v=0) rotational transition, and the lower peak is the HD line
of the $J=4-3$ (v=0) rotational transition.  The ratio of the peaks is
about 20, this value characterising the emission from primordial
galaxies.  The expected emission flux at the redshift of $1+z \sim 20$
(e.g. $\Omega_m = 0.3$ and $\Omega_\Lambda =0.7$), in the $J=2-0$
(v=0) line of H$_2$ occurs at a rate $\sim 2 \times 10^{-7}$ Jy, and
in the $J=4-3$ (v=0) line of HD at a rate $\sim 8 \times 10^{-9}$ Jy.
The former has a frequency of 5.33179$\times 10^{11}$ Hz and the
latter is at 5.33388 $\times 10^{11}$Hz, respectively.  Since the
frequency resolution of ALMA is about 40 kHz, the double peak is
resolvable.  While an individual object is not observable even by
ALMA, the expected assembly of primordial star clusters on subgalactic
scales can result in fluxes at the 2000-50 $\mu$Jy level. These are
marginally observable.  The first peak of H$_2$ is produced when the
core gas cools due to HD cooling, while the second peak of HD occurs
because the medium maintains thermal balance by H$_2$ cooling which
must be enhanced by three-body reactions to form H$_2$ itself.

\end{abstract}


\section{INTRODUCTION}

Observation of the first generation of stars presents one of the most
exciting challenges in astrophysics and cosmology.  Hydrogen molecules
(H$_2$s) play an important role as a cooling agent of the
gravitationally contracting primordial gas (Saslaw \& Zipoy 1967) in
the process of primordial star formation.  It has been argued that
H$_2$ is an effective coolant for the formation of globular clusters
(Peebles \& Dicke 1969), if such objects precede galaxy formation. The
contraction of primordial gas was also studied in the pioneering work
of Matsuda, Sato, \& Takeda (1969).  Thus, we expect the detection of
H$_2$ line emission from primordial star-forming regions (Shchekinov
1991; Kamaya \& Silk 2002; Ripamonti et al. 2002).  
Observational feedback from such Population III objects was
examined in Carr, Bond, \& Arnett (1984).  In the context of the CDM
(Cold Dark Matter) scenario for cosmic structure formation, Couchman
\& Rees (1986) suggested that the feedback from the first structures
is not negligible for e.g. the reionisation of the Universe and the
Jeans mass at this epoch (e.g. Haiman, Thoul, \& Loeb 1996; Gnedin \&
Ostriker 1997; Ferrara 1998; Susa \& Umemura 2000).  Recent progress
on the role of primordial stars formation in structure formation is
reviewed in Nishi et al. (1998) and its future strategy is discussed
in Silk (2000).

More recently, the first structures in the Universe have been studied
by means of very high resolution numerical simulations (Abel, Bryan,
\& Norman 2000).  The numerical resolution is sufficient to study the
formation of the first generation of molecular clouds. According to
their results, a molecular cloud emerges with a mass of $\sim 10^5$
solar masses as a result of the merging of small clumps which trace
the initial perturbations for cosmic structure formation. Due to the
cooling of H$_2$, a small and cold prestellar object appears inside
the primordial molecular cloud. It resembles the cores of molecular
clouds at the present epoch.  These numerical results are consistent
with other numerical simulations by Bromm, Coppi, \& Larson (1999;
2001).  All the simulations predict that the primordial molecular
clouds and their cores appear at an epoch of $1+z \sim 20$ ($z$ is
redshift).

According to these results, the first generation of young stellar
objects has a mass of $\sim$ 200 solar masses, density of $\sim 10^5$
cm$^{-3}$, and temperature of $\sim$ 200 K. We stress that these are
cloud cores, and not stars.  Inside the cores, a very dense structure
appears. We call this a {\it kernel} for clarity of presentation
(Kamaya \& Silk 2001).  Its density increases to a value as high as
$\sim 10^8$ cm$^{-3}$ where three-body reactions for H$_2$ formation
occurs (Palla, Salpeter, \& Stahler 1983, Omukai 2000).  
Further evolution is by fragmentation (Palla, Salpeter, \&
Stahler 1983) and/or collapse (Omukai \& Nishi 1998). In the previous
paper, we considered the collapsing kernel and the possibility for its
observation by future facilities.  H$_2$ line emission tracing the
temperature structure of the kernel is potentially detected by
ASTRO-F\footnote{http:// www.ir.isas.ac.jp/ASTRO-F/index-e.html} and
ALMA\footnote{http:// www.alma.nrao.edu} if the kernels are
collectively assembled, as might be expected in a starburst.
We also comment here that if there is dust grain in starforming region,
the situation changes (e.g. Hirashita, Hunt, Ferrara 2002).
For example, the three body reaction dominates over the grain surface 
formation if dust grain is present type and 
only when gas density is above $10^{11}$ cm$^{-3}$.

Now in the low temperature regime of the typical core, HD is also an
important coolant (e.g. Shchekinov 1986).  Indeed, Uehara \& Inutsuka
(1998) show numerically that the core collapses at a relatively low
temperature if HD exists. According to them, the primordial filament
can reach a temperature of 50 K. Thus, the effect of HD should be
examined. Indeed as long as HD is a dominant coolant, it is useful to
consider the observational possibility of HD emission from primordial
molecular cloud cores, as we do below.

Furthermore, because of the large electric dipole moment, LiH is also
potentially an important coolant (Lepp \& Shull 1984). Despite the
very low abundance of LiH, which means that it is never the dominant
coolant, as shown in Lepp \& Shull, if its lines are optically thin,
it can still be important.  We also comment on line cooling by LiH in
this paper.

In addition to a contracting primordial cloud, another molecular
emission line is discussed by Ciardi \& Ferrara (2001) who predict the
detection of mid-infrared emission from primordial supernova shell by
NGST\footnote{http://ngst.gsfc.nasa.gov}.  According to Ciardi \&
Ferrara, the shell can cool and radiate in rotational-vibrational
emission lines, and then mid-infrared emission is expected.  In
particular, the shell can also emit the possible ground rotational
line of H$_2$ ($J=2-0$ and $v=0$). In our first paper, we also predict
the primordial emission of the ground rotational line of H$_2$. If and
when primordial emission of the ground rotational transition is
detected, it will be important to judge which is its origin.  In this
paper, we present a very simple answer to this question.

In \S2, we formulate the cooling function for HD and LiH.  For H$_2$,
this was presented in a previous paper (Kamaya \& Silk 2001).  We
summarise the cooling function of H$_2$ in the Appendix.  In \S3, we
present a model structure for the primordial molecular cloud cores. In
\S4, we estimate that the luminosity from the cores is primarily due
to the emission of H$_2$, with a secondary role played by HD, and that
LiH is not important.  In \S5, the observational feasibility of
detection is reviewed (differently from the discussion in our previous
paper).  The theory underlying the observational predictions is
presented in \S6, and our paper is summarised in the final section.

\section{DESCRIPTION OF MOLECULAR EMISSION}

Primordial molecular cloud cores are expected to be low temperature
objects (Abel et al. 2000), where HD (Shchekinov 1986) and LiH (Lepp
\& Shull 1984) line cooling are frequently considered to be important.
In this section, first of all, we present a formalism for the
molecular line emission to estimate the luminosity from the cores.
The lower rotational transition emission of H$_2$ is also important.
Since a detailed description has been presented in the previous paper
(Kamaya \& Silk 2001), only a brief summary is given in the Appendix.

\subsection{HD}

HD has a weak electric dipole moment, since the proton in the molecule
is more mobile than the deuteron; the electron then does not follow
exactly the motion of the positive charge, producing a dipole (Combes
\& Pfenniger 1997). This moment has been measured in the ground
vibrational state from the intensity of the pure rotational spectrum
by Trefler \& Gush (1968). The measured value is about 5.85 $10^{-4}$
Debye (1 Debye $=$ $10^{-18}$ in cgs unit).  Since the first
rotational level is about 128 K, the corresponding wavelength is 112
$\mu $m.  Since the temperature of the primordial molecular cloud
cores (Abel et a. 2000) is about 200 K, HD has sufficient potential to
be the dominant coolant (Shchekinov 1986).  In this paper, we set the
D abundance to be $4\times 10^{-5}$, keeping consistency for our
structural model of cores with their thermal conditions predicted by
Uehara \& Inutsuka (2000), which is the most recent and detailed
analysis for a primordial cloud undergoing HD cooling.

For the cooling function, we adopt the formalism of 
Hollenbach \& McKee (1979). According to this, 
$$
\Lambda_{\rm HD,thin} = n \times n_{\rm HD} \times 
\frac{4 (kT)^2 A_0}{n E_0 ( 1 + (n_{\rm cr}/n)  
                              + 1.5 (n_{\rm cr}/n)^{0.5})}
~~~{\rm erg}~{\rm cm}^{-3}~{\rm s}^{-1}
\eqno(1)$$
where $k$ is the Boltzmann constant, $n$ is the gas number density,
$n_{\rm cr}$ is the critical density, $n_{\rm HD}$ is HD number density, 
$A_0$ is $5.12 \times 10^{-8}$ s$^{-1}$ (Abgrall et al. 1992),
and $E_0 = 64 k$ erg for HD.  Determining $ n_{\rm cr}$ exactly is a
complex calculation, while for our purpose it is sufficient to
consider $ n_{\rm cr}$ for the dominant cooling transition if we are
interested in the dominant process.  Hence, we estimate $n_{\rm cr}$
via the formalism of Hollenbach \& McKee (1979) as $n_{\rm cr} = 7.7
\times 10^3$ cm$^{-3}$ $(T_{\rm g}/1000.0)^{0.5}$ where $T_{\rm g}$ is 
the thermal temperature of gas. For reader's convenience, 
we would like to note that our $\Lambda_{\rm HD,thin}$ has changed 
from $L$ of (6.23) of Hollenbach \& McKee (1979) by the amount of
$n \times n_{\rm HD}$, while both are always consistent.

However, the above formula is inadequate if the temperature is low
since the population of the rotationally excited level is small.  We
consider the cooling function of HD in the optically thin and low
temperature regime below 255 K (Galli \& Palla 1998) if the density is
below the critical density for a given temperature;
$$
\Lambda_{\rm HD;n\to 0} =
2\gamma_{10}E_{10} {\rm exp}(-E_{10}/kT_{\rm g})
+(5/3)\gamma_{21}E_{21} {\rm exp}(-E_{21}/kT_{\rm g}).
\eqno(2)$$
Here, $E_{10} = 128k$ erg, $E_{21} = 255k$ erg ,
$\gamma_{10} = 4.4 10^{-12} + 3.6 10^{-13} T_{\rm g}^{0.77}$ 
cm$^3$ s$^{-1}$, and 
$\gamma_{21} = 4.1 10^{-12} + 2.1 10^{-13} T_{\rm g}^{0.92}$
cm$^3$ s$^{-1}$.
If the gas density is above the critical density, we simply estimate
the cooling function by two methods. The first is for the temperature
range from $128$ K to $255$ K. In this range, we simply use Eq.(1)
multiplied by exp($-256 {\rm K}/T_{\rm g}$). This is because the
population of $J=2$ to that of $J=1$ is reduced by the factor of
exp($256 {\rm K}/T_{\rm g}$).  The second is for if the temperature is
below 128 K.  In such low temperature case, as simple modification as
for the first method for correction is not applicable.  Fortunately,
the two-level system is a good approximation. Then, we estimate the
cooling function from the spontaneous de-excitation rate ($A$) of the
most probable rotational level of $J=1$ and the collisional
de-excitation rate.  The de-excitation rate is estimated from the
collisional cross section of de-excitation divided by a sound speed.
The collisional cross section of de-excitation is typically $2.0\times
10^{-16}$ cm$^2$ for HD (Schaefer 1990).  Our approximation breaks at
the outer edge of the core, where we smoothly extrapolate keeping the
thermal balance. Fortunately, this correction does not alter the main
conclusion and the dominant cooling level of $J$.

On the other hand, to find the cooling function in the optically
thick regime, we always need the Einstein A coefficient. We calculate
it approximately in the standard way. The coefficient is
$$ A_{J'J} = \frac{64\pi^4\nu_{J'J}^3}{3hc^3}D^2\frac{J}{2J+1}
\eqno(3)$$
where $\nu_{J'J}$ is the frequency of the $J'\to J $ transition, $h$
the Planck constant, $c$ the light speed, $D$ the electric dipole
moment, and $J$ the rotational quantum number.  The results via Eq.(3)
describe the exact values of Abgrall et al. (1982) for HD very well.

Once $J_{\rm d}(T(r))$ which is the dominant cooling level of $J$ is
determined at each position of the core, we can modify the optically
thin cooling function to the optically thick cooling function by
multiplication of the escape probability, $\epsilon$.  To obtain the
cooling function for the thick case, hence we define:
$$
\epsilon_{\nu_{J'J}} 
= \frac{1 - {\rm exp}(\tau_{\nu_{J'J}}) }{\tau_{\nu_{J'J}}}
\eqno(4)$$
and
$$
\tau_{\nu_{J'J}} = 
\frac{A_{J'J} c^3}{8 \pi \nu^3_{J'J}} 
  \left(\frac{g_{J'}}{g_J} - \frac{n_{J'}}{n_{J}}\right) n_J 
\frac{R_J}{\delta v}
\eqno(5)$$
where $R_J$ is the Jeans length, $\delta v$ is the velocity
dispersion, and $g_J$ the statistical weight of $2J+1$.  The velocity
dispersion corresponds to Doppler broadening, and is estimated to be
given by the sound speed. The inferred optically thick cooling
function is then
$$ \Lambda_{\rm HD,thick} \equiv 
\Lambda_{\rm HD,thin} \times   \epsilon_{\nu_{J'J}} .
\eqno(6)$$
The procedure of Jeans length shielding in (5) is useful for a  simple
analytical analysis (Low \& Lynden-Bell 1976; Silk 1977).

\subsection{LiH}

The LiH molecule has a much larger dipole moment of 5.9 Debye and the
first rotational level is only at 21 K.  Thus, although its abundance
is very small ($\sim 10^{-10}$; we adopt this value), there is a
plenty of possibility for LiH to be an important coolant (Lepp \&
Shull 1984).  For LiH, fortunately, the formalism of 
Hollenbach \& McKee (1979)
is a very good approximation. Then, we use Eq.(1)
with parameters for LiH in the thin regime, and the escape probability
correction is applied if the optical depth of the dominant line
emission is above unity. The adopted parameters are $A_0 = 0.0113$ s$^{-1}$
and $E_0/k = 21.0$ K, and $n_{\rm LiH}$ instead of $n_{\rm HD}$.  
For the corresponding de-excitation rates, we use the
results of detailed balance analysis (Bougleux \& Galli 1997). The
expected collisional de-excitation cross section is found to be
approximately $5.6 \times 10^{-16}$ cm$^{2}$, then the de-excitation
rate becomes $2\times 10^{-10}$ cm$^{3}$ s$^{-1}$ at $T_{\rm g}=3000$
K.  Since the de-excitation cross sections tend to be nearly constant
at low energies for collisions with neutral particles, the approximate
de-excitation rate at lower temperature can be obtained multiplied by
the factor of $(T_{\rm g}/3000 ~{\rm K})^{0.5}$. If the comparison
with the case of collision of He are interested, the reduced factor of
$3^{0.5}$ is adopted to account for the difference in the reduced mass
(Bougleux \& Galli 1997).  Their appendix B is useful for further
details.  If only the optically thin cooling function is needed,
appendix A.3 of Galli \& Palla (1998) is useful as long as the gas
density is below any critical density.

\section{STRUCTURAL MODEL OF PRIMORDIAL MOLECULAR CLOUD CORE}

According to recent numerical simulations, molecular cloud cores
appear prior to the formation of population III objects. Cores contain
a dense and cool kernel. In our previous paper, we consider how much
the kernel emits H$_2$ line luminosity. In the current paper, we
consider the case of the cores. Since the cores have lower temperature
than the kernels (e.g. Uehara \& Inutsuka 2000), HD and LiH must be
considered.

The emission properties are determined by the density and temperature
structure of the cores. Fortunately, a reasonably simple model is
possible according to Uehara \& Inutsuka (2000) and Omukai (2000). We
find a fitting formula for the distribution of H$_2$ and temperature
which is found by trial and error to approximately reach thermal
equilibrium (see the next section).  For $f_2(r)$ (solid line in
Fig.1);
$$ f_2(r) = 0.0001 + 0.495 \times 
\frac{ {\rm exp}\left( \frac{n(r)}{10^{11.0} {\rm cm^{-3}}} \right)
      -{\rm exp}\left( \frac{n(r)}{10^{11.0} {\rm cm^{-3}}} \right)}
     { {\rm exp}\left( \frac{n(r)}{10^{11.0} {\rm cm^{-3}}} \right)
      +{\rm exp}\left( \frac{n(r)}{10^{11.0} {\rm cm^{-3}}} \right)}
 \eqno(8) $$
where $n(r) = 10.0^8 {\rm cm^{-3}} (r/0.01~{\rm pc})^{-2.2}$ (the
maximum of $f_2(r)$ is set to be 0.5 by definition).  This describes
approximately the effect of the three body reaction to form H$_2$. For
$T(r)$ (dashed line in Fig.1);
$$ T(r) = 50 ~{\rm K} 
\left( \frac{n(r)}{10^{4.0} {\rm cm^{-3}}} \right)^{\frac{1}{5}}
. \eqno(9) $$ 
We hypothesise that all D and Li are in molecular form above a density
of $10^4$ cm$^{-3}$. This is partially supported by Uehara \& Inutsuka
(2000).

To examine the total emission energy, we need to determine the mass
distribution around the centre of the core, where the first star
emerges. Also, we assume a spherical configuration for the mass
distribution.  According to Omukai \& Nishi (1998), a high accretion
rate is realized if a similarity collapse occurs with the adiabatic
heat ratio of 1.1 (e.g. Suto \& Silk 1988).  The density distribution
is described as
$$ \frac{\partial {\rm ln} \rho (r)}{\partial {\rm ln} r} 
= \frac{-2}{2-\gamma} . 
\eqno(8)$$
Here, $r$ is the radial distance from the centre, $\rho (r)$ is the
mass density of atomic H, H$_2$ and He, and $\gamma $ is the specific
heat ratio.  We set the mass-density distribution of a
protostellar-core with $\gamma = 1.1$ as $\rho(r) = \rho_{\rm
0}(r/r_0)^{-2.2}$ where $\rho_{\rm 0}$ is 2.0$\times 10^{-20}$ g
cm$^{-3}$ and $r_0$ is 0.63 pc.  These values are appropriate for
fitting a typical protostellar core of Abel et al. (2000).

\section{MOLECULE EMISSION  LUMINOSITY}

We are interested in the dominant rotational emission, then we
calculate
$$J_{\rm max} = 
\frac
{\int_{\rm core} 4\pi r^2 \Lambda_{i, {\rm thick}}(r) J_{\rm d}(T(r)) dr}
{\int_{\rm core} 4\pi r^2 \Lambda_{i, {\rm thick}}(r) dr}. 
\eqno(10)$$
Here, for HD ($i$=HD)
$$
J_{\rm d}(T(r)) = J_{\rm d, HD}(T(r)) \equiv 
\left( \frac72 \times \frac{T(r)}{64.0~{\rm K}} \right)^{0.5}
,  \eqno(11)$$
and for LiH ($i$=LiH)
$$
J_{\rm d}(T(r)) = J_{\rm d, LiH}(T(r)) \equiv
 \left( \frac72 \times \frac{T(r)}{21.0~{\rm K}} \right)^{0.5}
,  \eqno(12)$$
and for H$_2$ ($i$=H$_2$)
$$
J_{\rm d}(T(r)) = J_{\rm d, H_2} (T(r)) \equiv
 \left( \frac72 \times \frac{T(r)}{85.0~{\rm K}} \right)^{0.5}
. \eqno(13)$$
Here, all $J_{\rm d}(T(r))$ mean the dominantly contributing $J$-level
to the statistical weight at temperature of $T(r)$ (Silk 1983).  Using
our fitting formula of $f_2(r)$ and $T(r)$, we find each $J_{\rm max}$
is about 4.0 for HD, 17.0 for LiH, and 2.3 for H$_2$,
respectively. Then, to estimate the dominant line luminosity of
$L_{\rm thick}$ of Eq.(6), we use $J=4$ for HD, $J=17$ for LiH, and
$J=2$ for H$_2$, respectively.  The exceptional treatment (but
reviewed in Kamaya \& Silk 2001) for H$_2$ is given in the Appendix,
according to which we can discuss $J=2-0$ transition of H$_2$
independent of any other transition.

The estimated total luminosity for a single core 
($\int_{\rm core} 4\pi r^2 \Lambda_{\rm thick} dr$), $L_{\rm single}$, is
$4.2 \times 10^{35}$ erg s$^{-1}$. 
Each  component is 
$4.1 \times 10^{35}$ erg s$^{-1}$ for H$_2$,  
$2.1 \times 10^{34}$ erg s$^{-1}$ for HD, 
and 
$6.9 \times 10^{30}$ erg s$^{-1}$ for LiH.
Here, first of all, we find that the contribution due to LiH is very
small.  This contradicts the conclusion of Lepp \& Shull (1984).
Fortunately, the reason is simple. Although Lepp \& Shull regarded all
LiH lines as optically thin, they are optically thick for our case in
the region where LiH would be important as suggested by Lepp \& Shull.

The line broadening is estimated to be 
$\sim \delta v_{\rm D} $
$= \left( {2k T}/{\mu (r) m_{\rm H}} \right)^{0.5} $
in the dimension of velocity. Here, $\mu (r)$ is mean molecular weight
at each position. Adopting this,
the luminosity per Hz (T=1000 K is assumed) is as  follows;  
 H$_2$ rotation emission of $J=2-0$ 
(1.1$\times 10^{13}$ Hz; 2.8$\times 10^{-2}$ mm) is 
4.0 $\times 10^{27}$ erg Hz$^{-1}$, 
HD rotation emission of $J=4-3$ 
(1.1$\times 10^{13}$ Hz; 2.8$\times 10^{-2}$ mm) is
2.0 $\times 10^{26}$ erg Hz$^{-1}$, 
and
LiH rotation emission of $J=17-16$ 
(7.4$\times 10^{12}$ Hz; 4.0$\times 10^{-2}$ mm) is 
9.8 $\times 10^{22}$ erg Hz$^{-1}$. 
Here, we consider the dominant $J_{\rm max}$ to $J_{\rm max}-1$
transitions for HD and LiH, and $J_{\rm max}$ to $J_{\rm max}-2$
transition for H$_2$.

The total cooling rate at each position is summarised in figure
2. Below $10^8$ cm$^{-3}$, HD is the dominant coolant, while H$_2$ is
dominant above this density.  This confirms the results of Lepp and
Shull (1984).  We also check if our model of the molecular cloud core
is reasonable or not. To do it, we estimate the heating rate.  In a
contracting cloud without dust, the compressional heating is generally
dominant. When the three-body reactions for H$_2$ formation occur,
chemical heating is also important.  We consider these two heating
mechanisms. For compressional heating, we estimate $c_s^2/t_{\rm ff}$
(e.g. Omukai 2000).  The specific free energy for the three-body
reaction is 4.48 eV.  Our result is displayed in figure 3.  The ratio
of cooling and heating is given in figure 4.  As clearly shown in this
figure, the deviation from the thermal balance is within a factor of
three.  Thus, our model structure for the core is consistent with
thermal balance between molecular line cooling and the expected
heating over all of the density range considered in this paper.

\section{OBSERVATIONAL FEASIBILITY OF DOUBLE PEAK EMISSION}

According to our first paper (Kamaya \& Silk 2001), some bright
emission lines from assembly of primordial young stellar objects
usually have sub-mJy flux as long as the redshift of $z$ is about
10--40. Then, when we are interested in the same range of the
redshift, it is sufficient to discuss only a typical case of
$z=19$. The expected emission of the same emission lines from
different redshifts can also have the similar flux level.  This
realises because the flux per frequency has apparently positive effect
of the redshift (e.g. Eq.(6) of Ciardi \& Ferrara 2001) and the sound
speed at the primordial starforming region should also be redshifted
(Toleman 1966, Kamaya \& Silk 2001).  
Furthermore, our estimate is reasonable if we
do not consider smaller $z$ than $\sim 6$ at which the reionisation of
the Universe occurs and the adopted assumptions for analysis break.

In the current paper, we need the redshifted
observational frequency and wavelength. For the three lines, we obtain 
0.53$\times 10^{12}$ Hz; 5.6$\times 10^{-1}$ mm for $J=2-0$ of H$_2$,
0.53$\times 10^{12}$ Hz; 5.6$\times 10^{-1}$ mm for $J=4-3$ of HD,
and
3.7$\times 10^{12}$ Hz; 8.1$\times 10^{-1}$ mm for $J=17-16$ of LiH.
The adopted $1+z$ is 20 according to the results of recent numerical
simulations. Obviously, the predicted frequency is located in the
range of ALMA.  Hence, we discuss the observational possibility of
detection by ALMA in the current paper. ALMA is a ground-based radio
interferometric facility, and will consist of many 12-m antennas.  A
detailed recent review is found in Takeuchi et al. (2001). According
to this, the 5$\sigma $ sensitivities at 350 $\mu$m, 450, 650, 850,
1.3 mm, 3.0 mm are expected to be 390, 220, 120, 16, 7.5, 4.6 $\mu$Jy,
respectively (8-GHz bandwidth).

The most prominent feature is predicted to be a double peak of HD and
H$_2$ emission. A schematic view is presented in figure 5, but it is
depicted at the coordinate of the core (i.e. they are not
redshifted). The most obvious feature is the difference of the peaks
of each line intensity. The ratio of H$_2$ to HD is about 20.
Detection of this double peak feature would confirms the presence of
primordial molecules in forming galaxies.  The difference in frequency
of both the molecules is about $10^8$ Hz.  Each of the precise values
is 5.3317 $10^{11}$ Hz for H$_2$ (J=2-0) and 5.3338 $10^{11}$ Hz for
HD (J=4-3), respectively.  The frequency difference is resolved
sufficiently by ALMA since it has frequency resolution of about $4
\times 10^4$ Hz.

To estimate the observed flux, we determine the distance to the
object. Then, we calculate it numerically from the following standard
formula:
$$ D_{19} = \frac{c}{H_0}
\int_0^{19} \frac{dz}{(\Omega_\Lambda + 
                       \Omega_{\rm M}(1+z)^{3.0})^{0.5}}
\eqno(14)$$
where $H_0$ is the Hubble parameter taken to be 75 km sec$^{-1}$
Mpc$^{-1}$, $\Omega_\Lambda$ is the cosmological constant parameter,
and $\Omega_{\rm M}$ is the density parameter.  We consider the case
$\Omega_\Lambda + \Omega_{\rm M} =1$, since our discussion bases on
the numerical results (e.g. Abel et al. 2000).  Adopting $D_{19}$, we
obtain the observed fluxes of each of the lines.  The results are
summarised in table 1.  Although the line-broadening is estimated from
$\nu_0 \delta v_{\rm D} /c$, we also correct it for the redshift
effect ($\nu_0$ is the central frequency).
For each of the parameter sets of ($\Omega_\Lambda,\Omega_{\rm M}$),
we obtain $D_{19} = 0.62 \times 10^{10}$ pc 
($\Omega_\Lambda=0,\Omega_{\rm M}=1$),
$D_{19} = 1.00\times 10^{10}$ pc
($\Omega_\Lambda=0.7,\Omega_{\rm M}=0.3$),
and 
$D_{19} = 4.58 \times 10^{10}$ pc
($\Omega_\Lambda=0.9,\Omega_{\rm M}=0.1$).
According to Table 1, the rotational line fluxes of $J=2-0$ (v=0) for
H$_2$ are 0.16 $\mu$Jy; 0.0083 $\mu$Jy for $J=4-3$ (v=0) for HD; and
0.000001 $\mu$Jy for $J=17-16$ (v=0) for LiH if $\Omega_{\rm M} +
\Omega_\Lambda =1$.  Thus, we conclude that a single core is not
easily observable even by ALMA.

However we note that if the cores are collectively assembled on a
sub-galactic scale (Shchekinov 1991), the agglomeration can be
detected by ALMA (Kamaya \& Silk 2001).  We shall estimate the number
of cores in a primordial galaxy.  Firstly, we must consider the
lifetime of a core able to show double peak emission.  The required
density and temperature distribution for such a core is possible when
the accretion rate of the contracting gas is about 0.01 $M_\odot $
year$^{-1}$ (Omukai \& Nishi 1998; Kamaya \& Silk 2001).  Then, the
life-time of the core with double peak emission is about $10^4$ years
as long as a massive star of $\sim 100 M_\odot$ forms inside the core.
A $10^6 M_\odot$ cloud (Abel et al. 2000) is expected to form 1000
such cores at a plausible efficiency of 0.1 in mass during its
life-time.  Taking its lifetime to be $\sim 10^5$ years as the
free-fall time of a $10^6 M_\odot$ cloud, we find that its luminosity
can reach 100 $L_{\rm single}$.

Next, we consider an entire primordial galaxy with $10^{11} M_\odot$.
It may form $10^9$ supernovae over its entire lifetime.  We assume
that it makes $10^7$ primordial massive stars, since this number of
massive stars gives enrichment to roughly 1 percent of the solar
metallicity.  If the burst of formation of primordial molecular cloud
cores occurs in the central region of the primordial galaxy (we
postulate 1 kpc as the size of the core-forming region), then there
are $10^4$ such giant molecular clouds with $10^6 M_\odot$ in the
core-forming region, as long as the massive star forms with a high
accretion rate of 0.01 $M_\odot$ year$^{-1}$ (Kamaya \& Silk 2001).
During this phase, the cumulative luminosity would be $10^6$ $L_{\rm
single}$.  However, the dynamical time-scale in the core-forming
region may be $\sim 10^7$ years (e.g. the duration of the starburst).
Then, we obtain $10^4~L_{\rm single}$ for the luminosity of the
primordial galaxy undergoing its first star formation burst,
conservatively assuming $\sim 10^5$ years as the life-time of giant
molecular clouds.  Finally, for molecular-line emitting
proto-galaxies, we obtain 1.6 mJy for $J=2-0$ of H$_2$, 0.83 mJy for
$J=4-3$ of HD and 0.00001 mJy for $J=17-16$ of LiH ($\Omega_{\rm M} =
0.3$ and $\Omega_\Lambda = 0.7$).  The estimated flux levels of H$_2$
and HD are consistent with Shchekinov (1991), in which only the lowest
transition lines for both the molecules were discussed. It may also be
better to say that these values are optimistic values.  In more
realistic conditions, furthermore, our simple time-dependent summation
scheme might break down.

According to the current status of the instrumentation for ALMA,
80-890 GHz is the allowed detectable range.  It is feasible for our
prediction of the redshifted emission.  Unfortunately, however, the
transmission is bad for the predicted feature at 530 GHz because of
the atmosphere of the Earth.  Then, it may be necessary to detect
H$_2$ and HD emission from a protogalaxy forming larger than $1+z=20$.
Since $\Omega_M < 1$, which seems to be reasonable if $\Omega_\Lambda
= 0.7$, the formation of the cores is prompted. Then, we can expect to
detect emission at a larger frequency than 530 GHz, and detection
should be feasible of the H$_2$ and HD double emission from the
assembly of the primordial molecular cores. Of course, we also expect
the emission later than $1+z=20$, since the Universe has been
reionised at $z \sim 6$. The latter case is also favourable for ALMA
because of the better sensitivity.

In the previous paper, we advocated a deep blank field survey.  Then,
in the current paper, we propose another observational strategy to
detect the double emission feature from the primordial molecular cloud
cores.  Measurement of the number counts of submillimeter sources is
planned as one key galaxy evolution project for ALMA (Takeuchi et
al. 2001).  According to the predictions, many submillimeter sources
are expected below sub-mJy levels.  Once the survey is operating, we
can utilise the submillimeter source count data.  Firstly, we pick up
the faint sources around QSOs since QSOs are expected at high density
peaks in the usual hierarchical model of cosmic structure formation
from primordial Gaussian-distributed density fluctuations. Primordial
galaxies are concentrated around QSOs. Secondly, we roughly check the
spectra.  Primordial galaxies do not have dust, while evolved systems
have dust. This means the continuum level of the flux at submillimeter
wavelengths is significantly different between primordial and evolved
galaxies. Unfortunately, it seems that the faint sources are observed
in the low frequency resolution mode according to Takeuchi et
al. (2001).  But, one can expect to find unresolved double peaks as a
bump in the observed spectral energy distribution, which is a
different feature from the blackbody emission by dust.  To find the
bump due to the double emission lines should not be a difficult task.
After this simple selection, we re-observe the candidates for
primordial submillimeter sources in the high resolution mode in
frequency. Finally, it is expected that the primordial cores can be
discovered showing the double peak spectral features due to H$_2$ and
HD, in which the ratio of the two peaks, which is about 50, confirms
the hypothesis of primordial emission.

Finally, we stress that we are able to recognise the difference
between H$_2$ emission from the assembly of primordial molecular
clouds and that supernova shells in primordial starforming regions
(Ciardi \& Ferrara 2001). According to figure 1 of Ciardi \& Ferrara,
the emission of the ground rotational line of H$_2$ $(v=0)$, which is
never the dominant cooling line because of the high temperature of the
shell, can have a flux level of sub-mJy to mJy at $z\sim $ 10. This
flux level is obtained via adoption of the sound velocity of 10 km
sec$^{-1}$ in the shell gas. Thus, both our predictions and those of
Ciardi \& Ferrara yield similar flux levels for the same molecular
line of H$_2$.  The difference between the two predictions is the HD
line to be associated with the H$_2$ line in our case.  To conclude,
if and when the double peak emission is found, it would strongly
indicate the primordial emission from the assembly of the first
starforming molecular clouds.

\section{THEORY OF DOUBLE PEAK EMISSION}

How is the double-peak emission produced?  We describe a theory for
double-peak emission from primordial molecular cloud cores.  In the
entire region of the core, the temperature is significantly lower than
515 K which corresponds to the transition energy of H$_2$
(J=2-0). Such low temperatures are possible only when HD cooling is
efficient. Then, the effect of the large volume of the low temperature
region lets the first and larger peak be a rotational emission line of
H$_2$ (J=2-0).  This is possible because of the thermalisation of the
gas as long as the gas temperature is above 100 K $(i.e. \sim 2^2
\times 85\times 2/7; $J=2$)$, since the rotational level of $J=2$ is
excited by chance.


The situation of the second peak is a little more complicated.  We
define a characteristic density for the three-body reaction to set in
as being $n_{\rm three} \sim 10^8$ cm$^{-3}$. Around the region of
this density, the cooling rates of HD and H$_2$ are comparable.  Once
it is confirmed that the temperature is determined by the balance of
the cooling of HD and H$_2$ with adiabatic heating, we find that a
temperature of $\sim 300$ K at a density of $n_{\rm three}$ is
realized. In other words, if the cooling of H$_2$ were to be
neglected, the temperature would be lower than 300 K since the
corresponding heating was not allowed there. By the way, since the
temperature of 300 K permits the thermal excitation of $J=4$ of HD,
then the rotational emission of HD (J=4-3) becomes possible. This
emission occurs since the gas temperature can reach 300 K because of
the H$_2$ cooling balancing the adiabatic heating of the
gravitationally contracting core.  This efficiency of H$_2$ cooling is
realized only when the three-body reaction among H atoms occurs. Thus,
the second peak is possible owing to the three-body reaction to form
H$_2$.

\section{SUMMARY}

One of the main goals of cosmology is to find the first generation of
stars. When they form, strong H$_2$ emission and weak HD emission is
expected as a double peak feature.  The weak HD emission is important
since it distinguishes between the H$_2$ emission from the primordial
molecular clouds and that from the primordial supernova shell (Ciardi
\& Ferrara 2001).  We have examined the observational feasibility of
the detection of the double peak emission. According to our analysis,
the double peak feature is marginally detectable by ALMA.  However,
the transmissivity of air for the expected typical emission is low for
the ALMA project as pointed out in our previous paper.  If future
telescopes are able to detect the double peak feature emission from
primordial molecular cloud cores, then either primordial cores and the
first stars have formed at different redshift from $z=19$.

\section*{ACKNOWLEDGEMENT}
H.K. is grateful to Profs. S.Inagaki, S.Mineshige, and Yu.Shchekinov 
for their encouragement. We have appreciated the referee's careful
reading very much.


\section*{APPENDIX}

We re-formulate the energy loss  due to line emission of H$_2$.  The
details are found in Kamaya \& Silk (2001).  Line emission of H$_2$
occurs due to the changes among rotation and vibration states.  We can
basically use the formulation of Hollenbach \& McKee (1979) for
rotational and vibrational emission of H$_2$.  Adopting their
notation, we get:
$$L_{\rm r} \equiv  
\left(
\frac{9.5\times 10^{-22} T_3^{3.76}}{1+0.12T_3^{2.1}} 
{\rm exp}\left[ -\left( \frac{0.13}{T_3}\right)^3 \right]
\right)
$$
$$ +
3.0 \times 10^{-24}{\rm exp}\left( -\frac{0.51}{T_3} \right)
~~~{\rm erg}~{\rm s}^{-1}, \eqno({\rm A}1)$$
then, we estimate the cooling rate as 
$$\Lambda ({\rm rot}) =
n_{\rm H_2} L_{\rm r} ( 1 + \zeta_{\rm Hr} )^{-1}
+ n_{\rm H_2} L_{\rm r} ( 1 + \zeta_{\rm H_2r} ) ^{-1}
~~~{\rm erg}~{\rm cm}^{-3}~{\rm s}^{-1} ,\eqno({\rm A}2)$$
where 
$\zeta_{\rm Hr} = n_{\rm Hcd}({\rm rot})/n_{\rm H}$,
$\zeta_{\rm H_2r} = n_{\rm H_2cd}({\rm rot})/n_{\rm H_2}$,
$n_{\rm Hcd}({\rm rot}) = A_J/\gamma_J^{\rm H}$, 
$n_{\rm H_2cd}({\rm rot}) = A_J/\gamma_J^{\rm H_2}$, and
$A_J$ is the Einstein $A$ value for the $J$ to $J-2$ transition;
$\gamma_J^{\rm H}$ is the collisional de-excitation rate coefficient
due to neutral hydrogen; and
$\gamma_J^{\rm H_2}$ is that due to molecular hydrogen.
The first term of $L_{\rm r}$ denotes the cooling coefficient due to 
the higher rotation level ($J>2$) and the second one due to
$J = 2 \to 0$ transition. The vibrational levels of both terms
are set to be zero.

For the vibrational transitions; 
$$L_{\rm v} =
6.7 \times 10^{-19} {\rm exp}\left[ -\left( \frac{5.86}{T_3}\right)^3 \right]
+
1.6 \times 10^{-18}{\rm exp}\left( -\frac{11.7}{T_3} \right)
~~{\rm erg}~{\rm s}^{-1}, \eqno({\rm A}3)$$
then, we get the cooling rate as being 
$$ \Lambda ({\rm vib}) = 
n_{\rm H_2} L_{\rm v} ( 1 + \zeta_{\rm Hv} )^{-1}
+ n_{\rm H_2} L_{\rm v} ( 1 + \zeta_{\rm H_2v} )^{-1}
~~{\rm erg}~{\rm cm}^{-3}~{\rm s}^{-1} ,\eqno({\rm A}4)$$
where 
$\zeta_{\rm Hv} = n_{\rm Hcd}({\rm vib})/n_{\rm H}$,
and
$\zeta_{\rm H_2v} = n_{\rm H_2cd}({\rm vib})/n_{\rm H_2}$.
Here, $n_{\rm Hcd}({\rm vib}) = A_{ij}/\gamma_{ij}^{\rm H}$, and
$n_{\rm H_2cd}({\rm vib}) = A_{ij}/\gamma_{ij}^{\rm H_2}$ where
$A_{ij}$ is the Einstein $A$ value for the $i$ to $j$ transition;
$\gamma_{ij}^{\rm H}$ is the collisional de-excitation rate coefficient
due to neutral hydrogen; and
$\gamma_{ij}^{\rm H_2}$ is that due to molecular hydrogen.
In our formula, only the levels of $v=0,1$ and 2 are considered.  This
is sufficient since the temperature is lower than 2000 K.  The first
term of $L_{\rm v}$ is a cooling coefficient of $\delta v =1$ and the
second term is that of $\delta v = 2$.  The second term has effect on
only the central region of our structural model, then it has no
significant contribution to our conclusion.  Combining Eq.(A2) and
Eq.(A4), we obtain the total cooling rate as $\Lambda^{\rm thin} =
\Lambda ({\rm rot}) + \Lambda ({\rm vib}) $ erg cm$^{-3}$ s$^{-1}$ in
the optically thin regime.  When we need a cooling function which can
be used in the optically thick regime, $\Lambda^{\rm thin}$ is
multiplied by the escape probability like the other cooling function
in the main text.

It may be better to comment on the continuum absorption.  The effect
of the continuum absorption below 2000 K would be multiplied by ${\rm
exp} (-\tau_{\rm cont})$, in which
$$\tau_{\rm cont} = \rho (r) \lambda_{\rm J} 
\left[
4.1\left(\frac1{\rho (r)}-\frac1\rho_0\right)^{-0.9} {T_3}^{-4.5} 
+
0.012\rho^{0.51}(r) T_3^{2.5} 
\right]
\eqno({\rm A}5)$$
according to the estimate of Lenzuni, Chernoff, \& Salpeter (1991) who
obtain a fitting formula for the Rossland mean opacity in a
zero-metallicity gas. Here, $\lambda_{\rm J}$ is the Jeans length and
$\rho_0$ is 0.8 g cm$^{-3}$.  Their fitting formula is reasonable if
we consider the temperature range $T>1000$ K.  Our lowest temperature
of the collapsing core is about 50 K. Then, their formula may not be
appropriate, while it gives an sufficiently upper limit if we adopt
$\tau_{\rm cont}$ of 1000 K instead of really having $\tau_{\rm cont}$
below 1000 K.  For all of our estimates, $\tau_{\rm cont}$ is much
smaller than unity. Then, we can neglect the effect of continuum
absorption.

We summarise the parameters in our calculation 
for a rotational transition with $v=0$; 
$A_{2,0} = 2.94 \times 10^{-11}$ sec$^{-1}$; and 
$A_{J,J-2} = 5A_{2,0}/162 \times J(J-1)(2J-1)^4/(2j+1) $
sec$^{-1}$ are considered. For a vibrational transition,
$A_{10} = 8.3 \times 10^{-7}$ sec$^{-1}$;
$A_{21} = 1.1 \times 10^{-6}$ sec$^{-1}$; and 
$A_{20} = 4.1 \times 10^{-7}$ sec$^{-1}$ are considered.
In the main text, 
we find that the rotational line cooling of H$_2$ is dominated
by $J=2-0$ transition ($v=0$).
Then, we regard that the second term of Eq.(A1) is important.
This means our estimate for $J=2-0$ transition is very reliable,
while the other estimate has some uncertainty.

Finally, we comment on the uncertainty of the cooling function owing
to rot-vibrational transitions of H$_2$.  The formula for the cooling
function of H$_2$ is examined by Martin et al. (1996), Forrey et
al. (1997), Galli \& Palla (1998), and Fuller \& Couchman (2000).
Fuller \& Couchman especially stress that there is uncertainty in the
H$_2$ cooling function because of the difficulty in calculating the
interaction potential at low temperatures.  Then, different choices
for rotational and vibrational H-H$_2$ rate coefficients will produce
differences in the cooling function.  Fortunately, we can consider our
cooling function to be applicable since the cooling rate via H$_2$
transitions balances the release of gravitational potential energy
consistently as shown in our first paper (Kamaya \& Silk 2001).


%
%

\clearpage

\begin{table}
\begin{center}
\caption{Expected Emission Lines} 
\begin{tabular}{crrrccc}
\hline \hline
--- & H$_2$    & HD         & LiH
    & $\Omega_m$ & $\Omega_\Lambda$  & $D_{19}$ ($10^{10}$ pc)\\
    & ($J=2-0$) & ($J=4-3$)  & ($J=17-16$) 
    &                 &            \\
\hline
$\nu$     (10$^{13}$ Hz; $z=0$) &1.06 &1.06 &0.74 & -- & -- & --\\
$\nu$     (10$^{12}$ Hz; $z=19$)&0.53 &0.53 &0.37  & -- & -- & --\\
$\lambda$ (10$^{-3}$ mm; $z=0$) &28.29 &28.29 &40.30 & -- & -- & --\\
$\lambda$ (10$^{-3}$ mm; $z=19$)&565.8 &565.8 &806.1& -- & -- & --\\
$L_\nu$   (10$^{26}$ erg sec$^{-1}$ Hz$^{-1}$) 
                       &40.1 &2.01  &0.001 &-- & -- & --  \\
$f_\nu$   (10$^{-8}$ Jy; $z=19$) 
                        &43.18 &2.16 & 0.0003& 1.0 & 0.0 & 0.62 \\ 
$f_\nu$   (10$^{-8}$ Jy; $z=19$)
                        &16.60 &0.83 & 0.0001& 0.3 & 0.7 & 1.00 \\
$f_\nu$   (10$^{-8}$ Jy; $z=19$)
                        & 7.58 &0.37 & 0.00004& 0.1 & 0.9 & 1.48 \\
\hline
\end{tabular}
\end{center}
\end{table}

\clearpage


\begin{figure}
\begin{center}
\rotatebox{270}
{\includegraphics[height=9cm,clip,keepaspectratio]{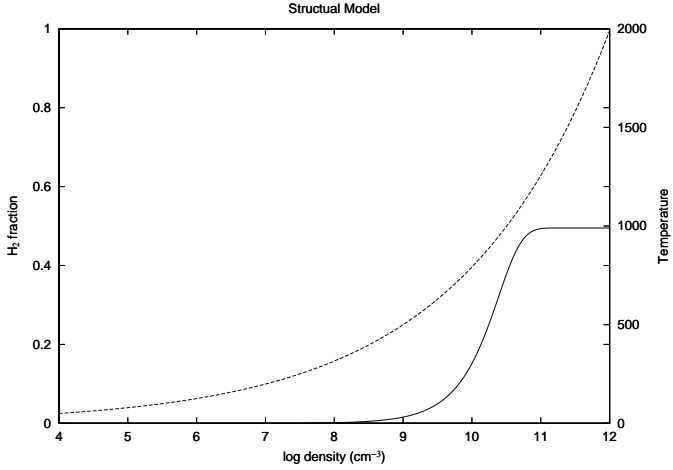}}
\caption{
Density--H$_2$ fraction relation (solid line) and
Density--Temperature relation (dashed line) are depicted.
Using Eq.(8),
we translate them to Radius--Density and --Temperature relations,
respectively.
}
\end{center}
\end{figure}

\begin{figure}
\begin{center} \rotatebox{270}
{\includegraphics[height=9cm,clip,keepaspectratio]{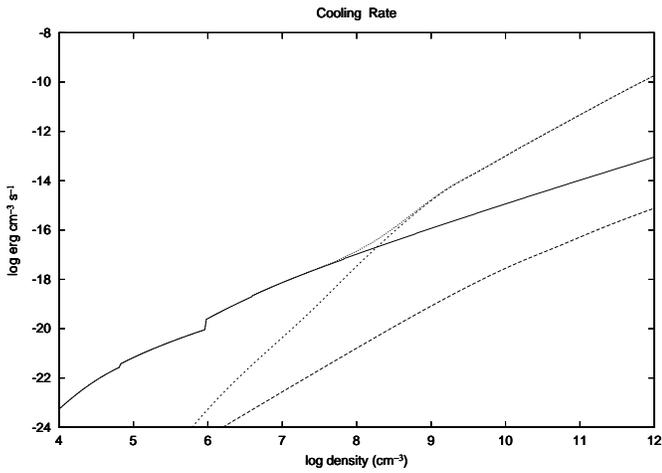}}
\caption{
Cooling rate; HD (solid), H$_2$ (short dashed), LiH (long dashed), and
total (dotted). Below $\sim 10^8$ cm$^{-3}$, HD is dominant,
while H$_2$ is important above the density. LiH is not 
so significant because of the lines of LiH is optically thick.
}
\end{center}
\end{figure}

\begin{figure}
\begin{center} \rotatebox{270}
{\includegraphics[height=8cm,clip,keepaspectratio]{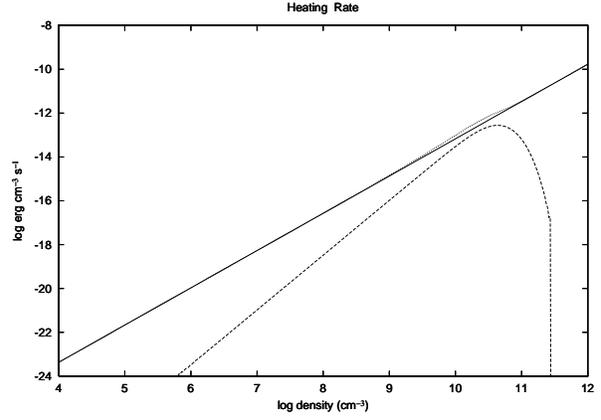}}
\caption{
Heating rate; compressional (solid), reactive (dashed), and
total (dotted).
}
\end{center}
\end{figure}

\begin{figure}
\begin{center} \rotatebox{270}
{\includegraphics[height=9cm,clip,keepaspectratio]{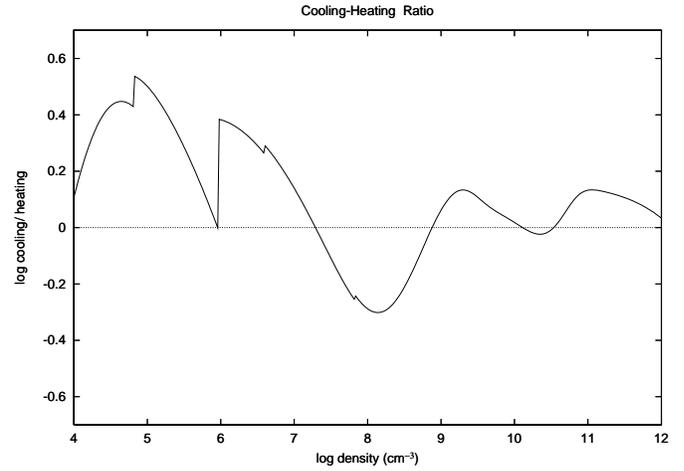}}
\caption{
Ratio of cooling to heating. As found clearly, our model
is reasonable almost a few factors. This supports our model
structure of a molecular cloud core. 
}
\end{center}
\end{figure}

\begin{figure}
\begin{center} \rotatebox{270}
{\includegraphics[height=9cm,clip,keepaspectratio]{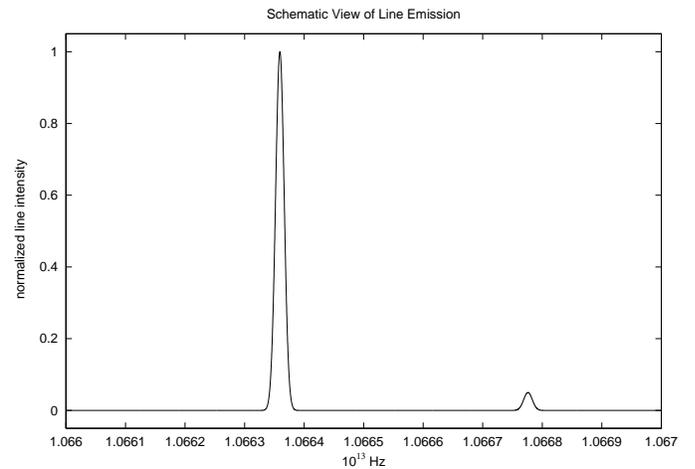}}
\caption{
Schematic view of the double peak emission of H$_2$ and HD.
The difference between the two lines is about 5.0$\times 10^{10}$ Hz.
This is resolved by ALMA project. 
}
\end{center}
\end{figure}

\end{document}